# TITLE
Determinant role of size-dependent electron relaxation processes in the nonlinear luminescence emission of resonantly excited gold nanorods

## AUTHORS
Céline Molinaro, Sylvie Marguet*, Ludovic Douillard, Fabrice Charra, Céline Fiorini-Debuisschert
SPEC, CEA, CNRS, Université Paris-Saclay, 91191 Gif-sur-Yvette cedex
*NIMBE, CEA, CNRS, Université Paris-Saclay, 91191 Gif-sur-Yvette, France

## ABSTRACT
The two-photon luminescence (TPL) of gold nanoparticles (NP) was shown to result from the excitation of hot carriers, the plasmonic NP resonances playing an important role both for plasmon enhanced absorption and plasmon enhanced emission. However, the exact parameters enabling to control or optimize the NP nonlinear luminescence still need to be understood in detail. In this paper, we report the two-photon excited photoluminescence of single gold nanorods exhibiting identical aspect ratio (close to 4) and thus identical plasmonic resonances, but increasing volumes V ($707 < V < 160\ 10^3$ nm$^3$ i.e. rod diameters varying between 6 and 40 nm). The two-photon luminescence intensity of a high number of colloidal nanorods was investigated at the single object level, combining polarization resolved TPL and simultaneously acquired topography. Non-monotonic TPL variations are evidenced, nanorods with an intermediate size (diameter around 10 nanometers) exhibiting the highest TPL signal intensity. A model is proposed considering both the local field enhancement effects at the NP and the size-dependent electron thermalization processes. BEM (Boundary Elements Method) simulations are used to compute the fields at both the transverse and longitudinal plasmon resonance. A good fitting of the experimental data is obtained considering integration of the fields over the whole the NP volume.

**Introduction**

Plasmonic luminescence has been the field of intense studies from applicative to more fundamental aspects. Beyond the on-going discussions regarding the origin of the signal,[1,2] two-photon luminescence (TPL) microscopy has proven to be a method of choice either for revealing the organization of plasmonic particles[3] or for mapping plasmonic hot spots and thus visualizing the local density of states inside metallic nanostructures.[4]

Conversely, if we consider e.g. the field of bioimaging for which plasmonic particles exhibit interesting characteristics as labels,[5,6] we may wonder about the parameters to play with in order to optimize TPL emission. How can TPL emission be correlated to the local field enhancement effects at a given plasmonic particle since TPL is only one specific relaxation channel after the excitation of plasmons[7,8]: this is the question underlying the work presented in this paper.

As already considered in previous studies,[9] the shape of plasmonic nanoparticle (NP) is one of the parameters that influence plasmonic luminescence. One difficulty however is the change of plasmon resonances depending on both the particle shape and size for a given metal.

The studies presented here are focused on gold nanorods (GNRs) since they are known to display one of the highest TPL efficiency.[9] The large TPL of GNRs was shown to result from an interplay between their two different localized surface plasmon resonances: $\lambda_{LSP}^T$ around 520/540 nm for the transverse resonance and $\lambda_{LSP}^L$ above 750 nm for the GNR dipolar longitudinal resonance, the latter depending on the GNR aspect ratio (AR).[10] Excitation of a GNR at its longitudinal plasmon frequency ($\lambda_{exc} \approx \lambda_{LSP}^L$) first lead to increased absorption and thus increased production of electron-hole pairs. Enhanced radiative recombination of these hot carriers is then observed through an antenna effect at the GNR transverse plasmon resonance.[11] Quite similarly, such "Purcell effect enhanced radiative recombination of hot carriers" was also observed for the one photon luminescence of gold nanorods.[7]

The next step consists now in determining the exact parameters enabling to control and optimize the radiative electron-hole recombination for a nanorod of given transverse and longitudinal resonances.

Above crystallinity which is known to have important effects onto plasmon damping and luminescence,[12] another parameter to consider is the NP volume, with retardation effects potentially limiting field enhancement for larger particles, while smaller particles might be affected by surface damping effects.

As further detailed in the following, we have studied the TPL of GNRs of different sizes, considering one given aspect ratio ($\approx$ 4) and thus almost non-varying transverse and longitudinal plasmonic resonances ($\lambda_{LSP}^T \approx 530$ nm and $\lambda_{LSP}^L \approx 800$ nm, respectively). Different experiments were performed considering different batches of crystalline plasmonic nanorods from colloidal chemistry. We focused our studies on NRs of 4 different diameters ranging from 6 to 40 nm with a nearly constant aspect ratio around 4. The TPL intensity of a quite large number of GNR of each batch was characterized at the single object level with the simultaneous acquisition of each GNR topography, as previously reported.[11] From one batch to another, quite large variations of the GNR TPL intensity could be observed, the highest TPL intensity being obtained in the case of the intermediate size GNRs (10 nm diameter GNRs).

A phenomenological model is proposed in order to tentatively explain such behavior, our purpose being to describe quantitatively the emission efficiency of nanorods of varying volume from simple considerations:
(1) local field enhancement effects, which were already proven to be the primary reason for increased TPL in coupled plasmonic dimers.[13,14]
(2) the influence of hot electron relaxation, the timescale of either electron-electron or electron-phonon processes governing plasmon relaxation from photon emission to thermal effects.[15,16]

Simulations using the Boundary Elements Method (MNPBEM[17]) were performed in order to simulate the field enhancement properties of each type of GNRs. A scaling law is proposed, taking into account that increased local field at the excitation wavelength leads increased hot electron generation through Landau damping. At the same time, increased local field at the emission wavelength leads to increased photonic density of states and thus increased radiation in the far field (Purcell effect).

As further detailed below, confrontation of the experimental data with the model shows that additional effects needs to be taken into consideration, mainly electron thermalization through Auger scattering.[18] Moreover, we show that TPL is resulting from field enhancement effects produced all over the GNR volume and not only at the surface.

# RESULTS AND DISCUSSION

Four colloidal GNR batches were investigated with diameters $d$ and length $L$ respectively ranging from $d$ = 6 nm to $d$ =39 nm, and $L$=27 nm to $L$=145 nm: GNR1 (d=6 nm, L=27 nm), GNR2 (d=10 nm, L =40 nm), GNR3 (d=21.5 nm, L=84 nm) and GNR4 (d=39 nm, L=145 nm). Their normalized extinction spectra, measured in aqueous solution, are displayed in figure SP1. As expected from their nearly constant aspect ratio, all 4 batches from GNR1 to GNR4 display nearly identical localized transverse and longitudinal plasmon resonances: $\lambda_{LSP}^T \approx 520$ nm and $\lambda_{LSP}^L \approx 800$ nm (809 nm, 790 nm, 830 nm and 811 nm for GNR1 to GNR4, respectively).

Complementary scanning electron microscopy (SEM) and transmission electron microscopy (TEM) characterizations were performed following immobilization of the different GNRs onto ITO coated glass cover slide (see figure SP2 in the Supplementary information), confirming their crystallinity together with their quite good polydispersity and purity.

**Polarization and exciting wavelength influence**

A large number (about 20) GNRs of each batch was studied at the single object level. Since the NP environment might affect the plasmon relaxation,[19,20] careful attention was paid to the sample preparation as described in the "Methods" section. As previously reported,[11,15] we made use of a set-up enabling the simultaneous measurement of each single GNR topography (AFM) and polarization resolved TPL intensity (see figure SP3).

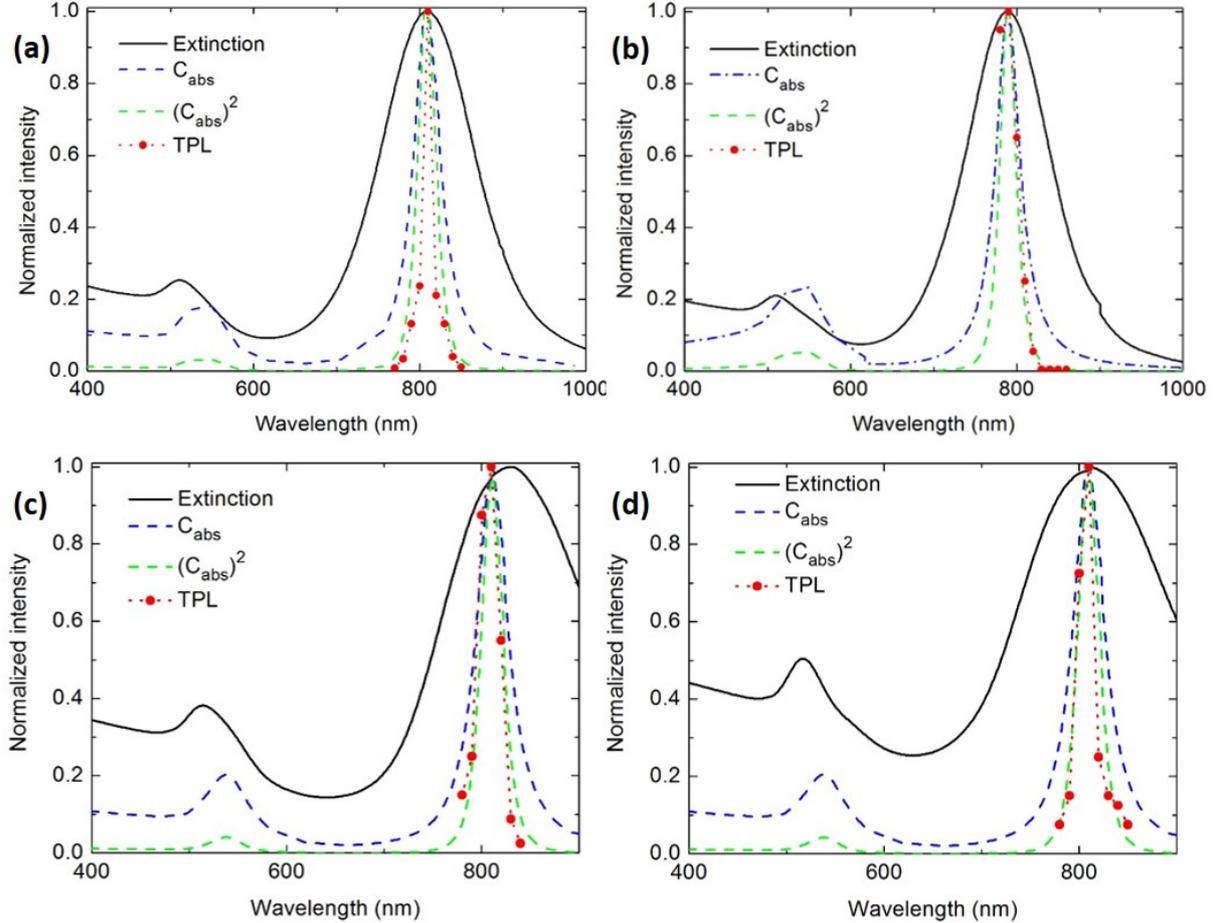

**Figure 1:** TPL excitation spectrum recorded for the individual GNRs of the different batches considered: (a) GNR1 (d=6 nm, L= 27 nm) ; (b) GNR2 (d=10 nm, L =40 nm) ; (c) GNR3 (d=21.5 nm, L=84 nm) and (d) GNR4 (d=39 nm, L =145 nm). Each measurement at varying excitation wavelengths is shown as a red disk, experimental curves being normalized to 1 (*see figure 2 further below for TPL signal comparison between each GNR batch*). For each batch, the extinction spectrum of a water solution (black line – ensemble measurements) of the corresponding GNR is shown for comparison. The dotted dashed blue and dashed green lines are the dependence of the GNR absorption cross section $C_{abs}$ and $C_{abs}^2$, respectively, as deduced from the Gans theory (normalized data).

The influence of both the polarization and wavelength of excitation was first studied showing features similar to what was previously observed in the case of 10 nm x 40 nm rods.[11] Whatever the considered batch, the highest TPL efficiency was observed in the case of an exciting polarization along the GNR longitudinal axis and for a resonant exciting wavelength $\lambda_{exc} = \lambda_{LSP}^L \approx 800$ nm. More particularly, as shown in figure 1, we observe that the TPL of each single GNR closely follows the wavelength dependence of its absorption: $I_{TPL}$ is directly proportional to $C_{abs}^2$, with $C_{abs}$ the GNR absorption cross section as calculated from the Gans theory.[21]

$$C_{abs} = \frac{2\pi}{3\lambda}\varepsilon_m^{3/2} V \sum_i \frac{\varepsilon_2/(n^{(i)})^2}{(\varepsilon_1+[(1-n^{(i)})/n^{(i)}]\varepsilon_m)^2+\varepsilon_2^2} \quad (1)$$

where $\varepsilon_m$ is the dielectric constant of the surrounding medium, $\varepsilon$ the dielectric constant of the metal with $\varepsilon = \varepsilon_1 + i\varepsilon_2$. $\lambda$ is the wavelength of the incident light and $n^{(i)}$ the depolarization factor described by:

$$n^{(a)} = \frac{2}{R^2-1}\left(\frac{R}{2\sqrt{R^2-1}}\ln\left(\frac{R+\sqrt{R^2-1}}{R-\sqrt{R^2-1}}\right) - 1\right) \quad (2)$$

$$n^{(b)} = n^{(c)} = (1 - n^{(a)})/2 \quad (3)$$

with R the GNR aspect ratio

**Spectral analysis of the TPL emission**

Whatever the considered GNR size, their TPL spectra were shown to be quite similar (see figure 2 below). The TPL consists in two emission bands peaking, one in the visible, very close to the GNRs transverse surface plasmon band ($\lambda_{TPL}^{max,VIS} \approx 540$ nm) and the second one in the IR, the latter band being cut due to the use of short pass filters to block the laser excitation.

An analysis of the TPL polarization properties could moreover show that TPL is mainly polarized along the short axis of the rods. This points to a preponderant radiative decay through the rod transverse plasmon mode, although "direct/unpolarized" radiative relaxation through interband relaxation near X and L symmetry points of the first Brillouin zone might also be possible.[22]

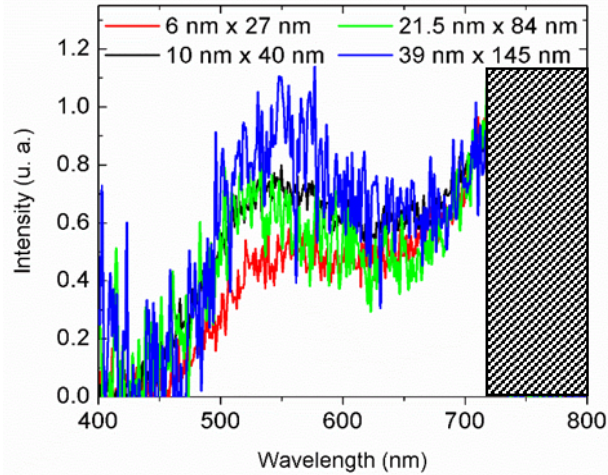

**Figure 2:** TPL spectrum of single GNRs: 6 nm x 27 nm (red), 10 nm x 40 nm (black), 21.5 nm x 84 nm (green) and 39 nm x 145 nm (blue). The hatched area corresponds to a non-measurable spectral range given the filters used to block laser excitation at the detector.

The TPL polarization rate was observed to decrease when considering GNRs of increasing volume (figure SP 4), in quite good agreement with the previously cited work in the group of Zhang, who have considered much bigger nanoparticles.[22]

**Influence of the GNR volume**

For each batch (given GNR volume), the TPL performances of about 20 individual GNRs were carefully analyzed, checking both the wavelength and polarization to get the maximum TPL intensity. The average excitation power was constant ~20 µW, i.e. much below the GNRs photo-induced damage threshold.[15] The compiled results are reported in figure 3.

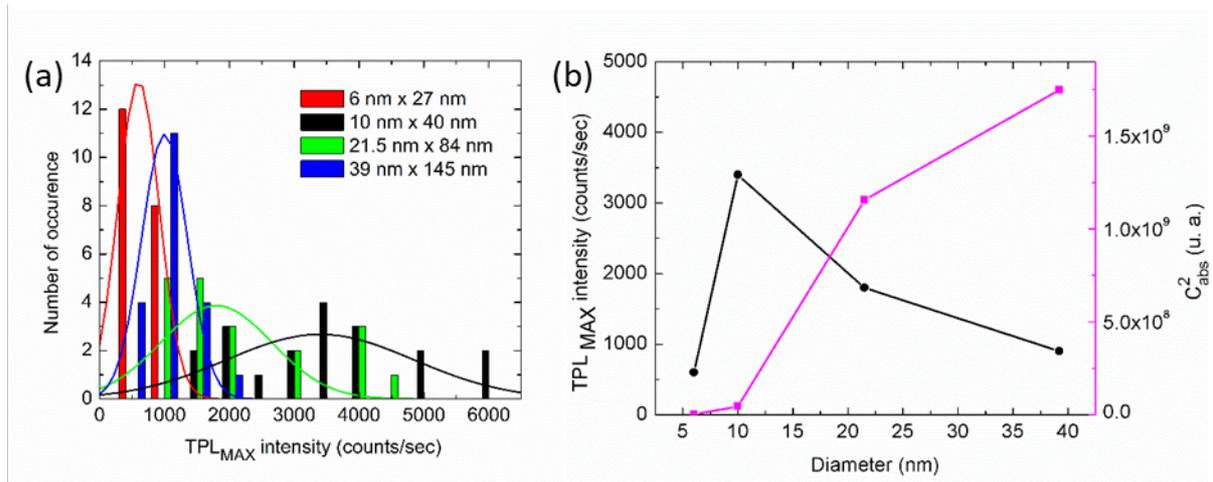

**Figure 3:** (a) Distribution in TPL intensity for each GNRs batch. Experimental data histograms (bars) are each fitted with a gaussian profile (solid lines): GNRs with d=6 nm (red), d=10 nm (black), d=21.5 nm (green) and d=39 nm (blue). (b) : For each batch (associated to a given diameter), the median TPL intensity is reported with a black line connection as a guide to the eyes. Similarly, the absorption cross-section $C_{abs}^2$ calculated from the Gans theory is plotted in pink.

For each considered GNR volume, the distribution in TPL intensity recorded for all the GNRs tested in each batch was adjusted with a Gaussian function, as represented in figure 3a. Quite large variations of the median value are clearly evidenced from one batch to another. Whatever the considered batch, half width at the maximum corresponds roughly to 40% of the median value, signing comparable polydispersity in each batch. The batch with the smallest GNR presented the lowest TPL intensity, with a median value of 600 counts/sec. The batch with d=21.5 nm x L=84 nm displayed a lower median value of 1800 counts/sec. Finally, the d=39 nm x L=145 nm GNRs showed an even lower TPL intensity with a median value of 900 counts/sec. The strongest TPL intensity (median value of 3 400 counts/sec) was clearly obtained for the intermediary volume of the d=10 nm x L=40 nm GNR.

Although the GNR absorption cross section $C_{abs}$ can explain the spectral dependence of TPL (figure 1), there happen thus to be no direct quantitative link between $I_{TPL}$ and $C_{abs}$, as evidenced in figure 2b. Absorption is indeed only one step in TPL. As further detailed below, the observed non monotonic dependence of TPL intensity with the GNRs volume requires to go deeper into the process and to consider all the phenomena involved in the luminescence of plasmonic nanoparticles.

**Scaling law – Correlation with BEM simulations**

In the case of one-photon luminescence, although former experiments could not necessarily highlight such non-monotonic behavior with size,[18,23] a microscopic theory for plasmon-enhanced was proposed by Shabazyan showing the emergence of an optimal NP size of 60 nm, considering spherical gold particles.[24,25]

The case of TPL in GNRs is however more complex requiring to consider two different plasmon resonances, since excitation is performed at the GNRs longitudinal plasmon while luminescence emission occurs at their transverse plasmon mode.

As discussed above, from a phenomenological point of view, two-photon luminescence can be described as a 3 step process:

(1) The first step is the enhanced production of hot carriers pairs $n_{hc}$ following a two-photon absorption process from the d-band to unoccupied states in the sp-band. Enhanced two-photon absorption results from an increased local exciting field when excitation is performed at the longitudinal resonance, leaving : $n_{hc} \propto f_{loc,\omega}^4$, with $f_{loc,\omega}$, the local field factor at the longitudinal resonance frequency ($f_{loc,\omega} = E_{loc,\omega}/E_{inc,\omega}$ with $E_{loc,\omega}$ the local electric field at $\omega$ frequency and $E_{inc,\omega}$ the incident electric field at $\omega$).

(2) The second step is a thermalization process during which the hot carriers relax towards the Fermi level, through electron-electron Auger scattering processes finally leading to the excitation of transverse plasmons. For gold spherical nanoparticles of radius $r$, the efficiency $\eta_{Auger}$ of such process scales as : $\eta_{Auger} \propto 1/r^3$ [18]. In order to adapt this formula to GNRs, we propose to use the effective radius $r_{eff}$ that can be calculated from the GNR volume : $V = \frac{4}{3}\pi r_{eff}^3$

(3) The last step is the radiation of this transverse plasmon, with efficiency $\eta_{rad}$ : the increased local density of states at the GNR transverse resonance leads to increased radiation in the far field.[7] The local density of states is directly dependent on the local field enhancement at resonance, leaving : $\eta_{rad} \propto f_{loc,\omega TPL}^2$, with $f_{loc,\omega TPL}$, the local field at the TPL emission frequency.

Numerical simulations resulting from the Boundary Elements Method (BEM)[17] were implemented in order to evaluate the local field factors at both the GNRs transverse and longitudinal resonance. Four GNRs were considered with sizes close to the experimental samples.

All presented simulations were performed by solving the full Maxwell's equations (retarded fields) as further detailed in the "Materials and Methods" section below. For each GNR, the field enhancement maps were computed at both the longitudinal and the transverse resonance. The particular case of the 10 nm x 40 nm GNRs is shown in figure 4(a,b). Figures SP5 and SP6 display the case of all the considered GNRs. As a complement, figure 4c and 4d provide cross sections of the electromagnetic field maps along the long and short axis of each GNRs at the longitudinal and transverse resonance, respectively. Given the symmetry of the rods and their homothetic sizes, this provides a direct comparison of the local electric fields $f_{loc,\omega}$ and $f_{loc,\omega TPL}$ at the excitation or the TPL emission frequency, respectively.

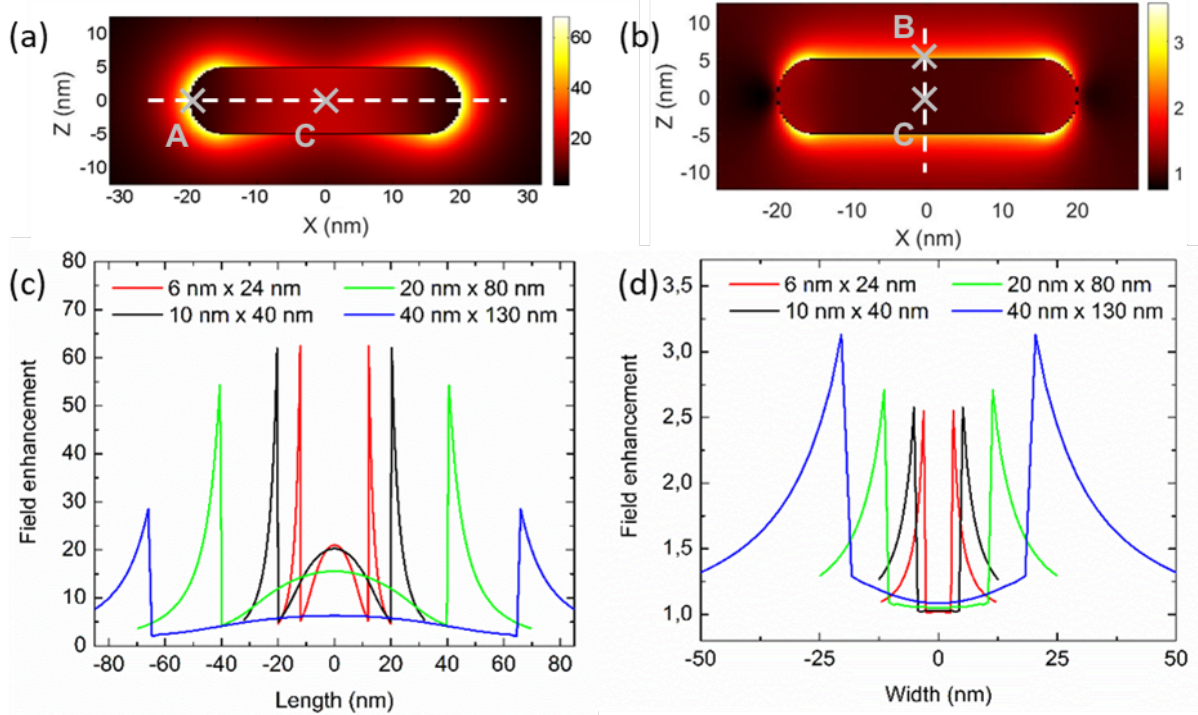

**Figure 4:** BEM simulations results. Calculated local field factors in intensity at the excitation ((a), $f^2_{loc,\omega exc}$, field polarization along the GNRs long axis) and emission wavelengths ((b), $f^2_{loc,\omega TPL}$, field polarization along the GNRs short axis) for a 10 nm x 40 nm GNR. The considered wavelengths are 799 nm ($\omega_{exc}$) and 522 nm ($\omega_{TPL}$), respectively corresponding to the calculated resonances (maximum absorption at $\omega_{exc}$ and maximum scattering at

ω<sub>TPL</sub>), in very good agreement with the experimentally determined values. The dashed white lines materialize the cross-sections along which the curves represented in (c) and (d) were obtained. For the sake of clarity each GNR size is associated to a specific color: GNR 6 nm x 24 nm (red line), 10 nm x 40 nm (black line), 20 nm x 80 nm (green line) and 40 nm x 130 nm (blue line).

At the excitation wavelength, we observe that both the internal and external field enhancement factors decrease with increasing the GNR volume. This is due to retardation effects with increased dephasing at increased GNR sizes.[26] At the emission wavelength, both field enhancement factors decrease with decreasing the GNR volume, although differences are quite limited.

From the discussion above, the luminescence intensity can then be calculated following integration over the whole particle:

$$I_{TPL} \propto \int n_{hc} \cdot \eta_{therm} \, \eta_{rad} \propto \frac{1}{r_{eff}^3} \int f_{loc,\omega_{exc}}^4 f_{loc,\omega_{TPL}}^2 \quad (4)$$

With $n_{hc}$ the number of hot carriers produced at a given point of the GNR, following the longitudinal plasmon excitation; $\eta_{Auger}$, the Auger thermalization rate and $\eta_{rad}$, the efficiency of radiative plasmon emission at the transverse resonance. $f_{loc,\omega}$ and $f_{loc,\omega TPL}$ stand for the local field factors at the excitation or the TPL emission frequency, respectively. They can be directly deduced from the BEM simulations. $\int$ is standing for integration over the particle, considering 2 different cases : full volume or surface integration.

Considering that GNRs present homothetic shapes, a simple scaling law might be considered to evaluate the variations in I<sub>TPL</sub> from one GNR to another, which enables to avoid any complicated integration calculations. Equation 4 can then be rewritten as equations (5) and (6) below considering integration over either the whole particle volume $I_{TPL}^{(V)}$ or surface $I_{TPL}^{(S)}$:

$$I_{TPL,V} \propto V \frac{1}{r_{eff}^3} f_{int,\omega_{exc}}^4(C) f_{int,\omega_{TPL}}^2(C) \, C_V \quad (5)$$

where $f_{int}$ is the value of the internal field enhancement taken at the center of the particle (C) and $C_V$ is a constant taking into account integration over the whole elliptic GNR volume.

$$I_{TPL,S} \propto S \frac{1}{r_{eff}^3} f_{ext,\omega_{exc}}^4(A) f_{ext,\omega_{TPL}}^2(B) C_S \quad (6)$$

where $f_{ext}$ is the value of the external field enhancement taken at the NP boundaries (apex A regarding the field at ω<sub>exc</sub> and place denoted B along the NP diameter as denoted in figure 4b) and $C_S$ is a constant taking into account integration over the whole elliptic GNR surface.

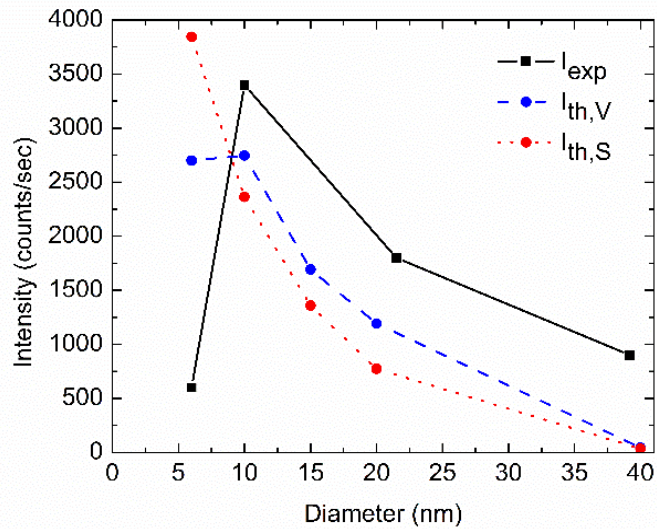

**Figure 5:** Experimental TPL intensity data (dark squares) and simulated TPL intensity resulting from the proposed model, considering either integration over the whole particle surface ($I_{TPL,V}$, red dots) or a bulk process ($I_{TPL,V}$, blue dots). An extra point corresponding to a GNR 15 nm x 60 nm was calculated (see figure SP9 for detailed BEM calculations).

The results are plotted in figure 5. The calculations considering the whole particle volume are in quite good agreement with the experimental results, confirming the validity of the proposed phenomenological model and evidencing moreover that TPL occurs mainly from the GNR bulk. A slight discrepancy can however be observed in the case of the smallest GNRs, for which the experimental TPL appears quite below the calculated value. It might be explained from surface induced losses which might be preponderant for very small objects.[27]

# CONCLUSION

The TPL intensity of four batches of GNR with different volumes and close longitudinal resonance wavelengths was investigated at the single objects level. From batch to batch, fairly large variations in the GNR TPL intensity of could be observed, and a non-monotonic dependence of TPL emission with GNR sizes was observed. In particular, the highest TPL intensity was observed for mid-size GNRs (d=10 nm, L=40 nm).

A phenomenological model was proposed taking into account the different processes involved in TPL from local field enhancement effects at the GNRs longitudinal and transverse plasmon resonances, to hot electrons relaxation.

Implementation of this model using BEM simulations was shown to explain quantitatively the measured TPL emission of the considered nanorods. Quite interestingly, this simple model enables to make the difference between either surface or volume contributions in the TPL emission. A good agreement could be obtained between the experimental results and the model only when considering the whole particle volume. TPL is thus concluded to be mostly a bulk process although specific features might sometimes be observed from surface defaults in the case of non-crystalline objects as recently reported in the litterature.[28]

Resonantly excited plasmonic nanoparticles (NP) are also known to activate chemical transformations directly in their surroundings, from energy conversion[29] to photocatalysis,[30,31] phototherapy or nanofabrication.[32] From the results presented in this article, we believe that TPL should be a valuable method to quantitatively transduce any changes in the particle surroundings and could thus be taken into profit to monitor the plasmon mediated chemical activity of metallic particles. Luminescence being only one specific relaxation path for plasmons, any other competing process such as hot carriers transfer shall also lead to a reduction of luminescence. For this, the phenomenological model proposed in this article will have to be further extended.

# METHODS

**Samples preparation**
GNRs, 6 nm x 27 nm and 10 nm x 40 nm, were synthesized using a seed-mediated procedure.[33,34] GNRs, 21.5 nm x 84 nm and 39 nm x 145 nm, also obtained from colloidal chemistry, were purchased from Nanopartz™ (*NANOPARTZ A12-25-808-CTAB* and *NANOPARTZ A12-40-808-CTAB*). Careful attention is paid to sample preparation since any surface defects at the particles might change their properties.[28] For each batch of GNRs, the same protocol is applied for sample preparation. A droplet of a dilute water solution of GNRs is left to dry onto Indium Tin Oxide (ITO) coated glass cover slides, after a previous UV ozone treatment. This treatment produces hydroxyl groups at the ITO surface, enabling an electrostatic interaction with the GNRs positively charged Cetyl Trimethyl Ammonium Bromide (CTAB) capping layer. Then the substrate is rinsed thoroughly with distilled water and ethanol to eliminate CTAB in excess or non-immobilized GNRs. After deposition, the different samples are characterized using scanning electron microscopy (SEM). Electron beam lithography landmarks enable the identification of specific objects, making it possible to perform complementary studies considering the same object, from SEM to AFM and TPL.

**Experimental set-up**
The experimental design is the same as already presented.[11,15] The two-photon-luminescence of single GNRs was studied using an inverted microscope set-up associated to a femtosecond Ti-sapphire laser excitation (Tsunami, Spectra Physics, 760 nm < $\lambda$ < 930 nm) and coupled to an Atomic Force Microscope (NanoWizard III, JPK). The incident light is focused on different single GNRs using an oil-immersion microscope objective (100X, NA = 1.3). The emitted light is collected through the same microscope objective and separated from the incident light using a dichroic mirror. In order to avoid the incident exciting laser beam to be detected, low-pass infrared filters are set just after the microscope output. The two-photon signal is analyzed using a channel-plate multiplier (Perkin-Elmer) and a spectrometer associated to a camera (Andor DU 401-BR-DD).

**BEM simulations**
Numerical simulations based on the boundary element method (BEM) are performed using the multiple nanoparticles BEM (MNPBEM14) Matlab toolbox.[17] The BEM solve Maxwell's equation, with the surface integral equation method[35] dealing with surface currents and charges. Only the surface need to be discretized. More details about the method can be found in the book of W.L. Wendland[36].
GNR are modeled by a cylinder associated to 2 hemispheres at each end. The refractive index of gold is extracted from the Johnson and Christy gold table.[37] Simulations are carried out with a single GNR in a medium with a constant refractive index of water $n$ = 1.33. Indeed as mentioned in our previous studies[11] the refractive index of water happens to be similar to the effective refractive index of a GNR immobilized at an ITO-air interface (n = (1+1.6)/2=1.3 where 1 and 1.6 are the refractive index of air and ITO respectively). Before mapping the field enhancement, extinction, scattering and absorption spectrum were calculated in order to get the resonance excitation wavelength of each GNR. Mapping was performed at the two GNR plasmon resonance: the excitation polarization is set along the long axis of the rod for the excitation map at longitudinal resonance and along the short axis of the rod for the emission map, at the transverse resonance. All simulations are taking into account retardation effects.

*Acknowledgements*. The authors would like to thank the C'nano 2010 program "INPACT", ANR Blanc 2012 "HAPPLE" and ANR PNANO 2013 "SAMIRé" projects. Jeremie BEAL (University of

Technology, Troyes) and Wajdi Heni (IS2M, Mulhouse) are gratefully acknowledged for the realization of landmarks using e-beam lithography and high resolution SEM imaging, respectively.*Supporting Information available:*
This document contains additional figures concerning the characterization of the GNRs (figures SP1 and SP2), the measurement of the polarization rate of TPL emission (Table 1), the description of the experimental set-up (figure SP3), and the BEM simulations in retarded mode for each batch of GNR considered (figures SP5 to SP9).

Technology, Troyes) and Wajdi Heni (IS2M, Mulhouse) are gratefully acknowledged for the realization of landmarks using e-beam lithography and high resolution SEM imaging, respectively.

*Supporting Information available:*
This document contains additional figures concerning the characterization of the GNRs (figures SP1 and SP2), the measurement of the polarization rate of TPL emission (Table 1), the description of the experimental set-up (figure SP3), and the BEM simulations in retarded mode for each batch of GNR considered (figures SP5 to SP9).

# REFERENCES

(1) Roloff, L.; Klemm, P.; Gronwald, I.; Huber, R.; Lupton, J. M.; Bange, S. Light Emission from Gold Nanoparticles under Ultrafast Near-Infrared Excitation: Thermal Radiation, Inelastic Light Scattering, or Multiphoton Luminescence? *Nano Lett.* **2017**, *17* (12), 7914–7919. https://doi.org/10.1021/acs.nanolett.7b04266.

(2) Cai, Y.-Y.; Sung, E.; Zhang, R.; Tauzin, L. J.; Liu, J. G.; Ostovar, B.; Zhang, Y.; Chang, W.-S.; Nordlander, P.; Link, S. Anti-Stokes Emission from Hot Carriers in Gold Nanorods. *Nano Lett.* **2019**, *19* (2), 1067–1073. https://doi.org/10.1021/acs.nanolett.8b04359.

(3) Rožič, B.; Fresnais, J.; Molinaro, C.; Calixte, J.; Lau-Truong, S.; Félidj, N.; Kraus, T.; Dupuis, V.; Hegmann, T.; Fiorini-Debuisschert, C.; et al. Oriented Single Gold Nanorods and Gold Nanorod Chains within Smectic Liquid Crystal Topological Defects. *Submitted*.

(4) Viarbitskaya, S.; Teulle, A.; Marty, R.; Sharma, J.; Girard, C.; Arbouet, A.; Dujardin, E. Tailoring and Imaging the Plasmonic Local Density of States in Crystalline Nanoprisms. *Nat. Mater.* **2013**, *12* (5), 426–432. https://doi.org/10.1038/nmat3581.

(5) Sreenivasan, V. K. A.; Zvyagin, A. V.; Goldys, E. M. Luminescent Nanoparticles and Their Applications in the Life Sciences. *J. Phys. Condens. Matter* **2013**, *25* (19), 194101. https://doi.org/10.1088/0953-8984/25/19/194101.

(6) Xin, H.; Namgung, B.; Lee, L. P. Nanoplasmonic Optical Antennas for Life Sciences and Medicine. *Nat. Rev. Mater.* **2018**, *3* (8), 228. https://doi.org/10.1038/s41578-018-0033-8.

(7) Cai, Y.-Y.; Liu, J. G.; Tauzin, L. J.; Huang, D.; Sung, E.; Zhang, H.; Joplin, A.; Chang, W.-S.; Nordlander, P.; Link, S. Photoluminescence of Gold Nanorods: Purcell Effect Enhanced Emission from Hot Carriers. *Acs Nano* **2018**, *12* (2), 976–985. https://doi.org/10.1021/acsnano.7b07402.

(8) Zhang, Y.; Nelson, T.; Tretiak, S.; Guo, H.; Schatz, G. C. Plasmonic Hot-Carrier-Mediated Tunable Photochemical Reactions. *ACS Nano* **2018**, *12* (8), 8415–8422. https://doi.org/10.1021/acsnano.8b03830.

(9) Gao, N.; Chen, Y.; Li, L.; Guan, Z.; Zhao, T.; Zhou, N.; Yuan, P.; Yao, S. Q.; Xu, Q.-H. Shape-Dependent Two-Photon Photoluminescence of Single Gold Nanoparticles. *J. Phys. Chem. C* **2014**, *118* (25), 13904–13911. https://doi.org/10.1021/jp502038v.

(10) Ni, W.; Chen, H.; Su, J.; Sun, Z.; Wang, J.; Wu, H. Effects of Dyes, Gold Nanocrystals, PH, and Metal Ions on Plasmonic and Molecular Resonance Coupling. *J. Am. Chem. Soc.* **2010**, *132* (13), 4806–4814. https://doi.org/10.1021/ja910239b.

(11) Molinaro, C.; El Harfouch, Y.; Palleau, E.; Eloi, F.; Marguet, S.; Douillard, L.; Charra, F.; Fiorini-Debuisschert, C. Two-Photon Luminescence of Single Colloidal Gold Nanorods: Revealing the Origin of Plasmon Relaxation in Small Nanocrystals. *J. Phys. Chem. C* **2016**, *120* (40), 23136–23143. https://doi.org/10.1021/acs.jpcc.6b07498.

# Supplementary Information

## 1 – Extinction spectra of the rods

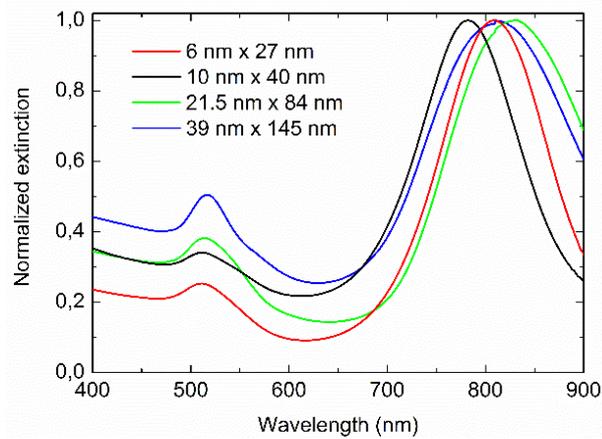

**Figure SP1:** Normalized extinction spectra of GNRs with respective diameters (d) and length (L): GNR1, d·L = 6 nm x 27 nm (red line, home-made particles see *Methods*); GNR2 : d·L = 10 nm x 40 nm (black line, home made particles, see *Methods*); GNR3, d·L = 21.5 nm x 84 nm (green line, *NANOPARTZ A12-25-808-CTAB*) and GNR4, d·L = 39 nm x 145 nm (blue line, *NANOPARTZ A12-40-808-CTAB*) in aqueous solution.

## 2 - SEM Characterization of the rods

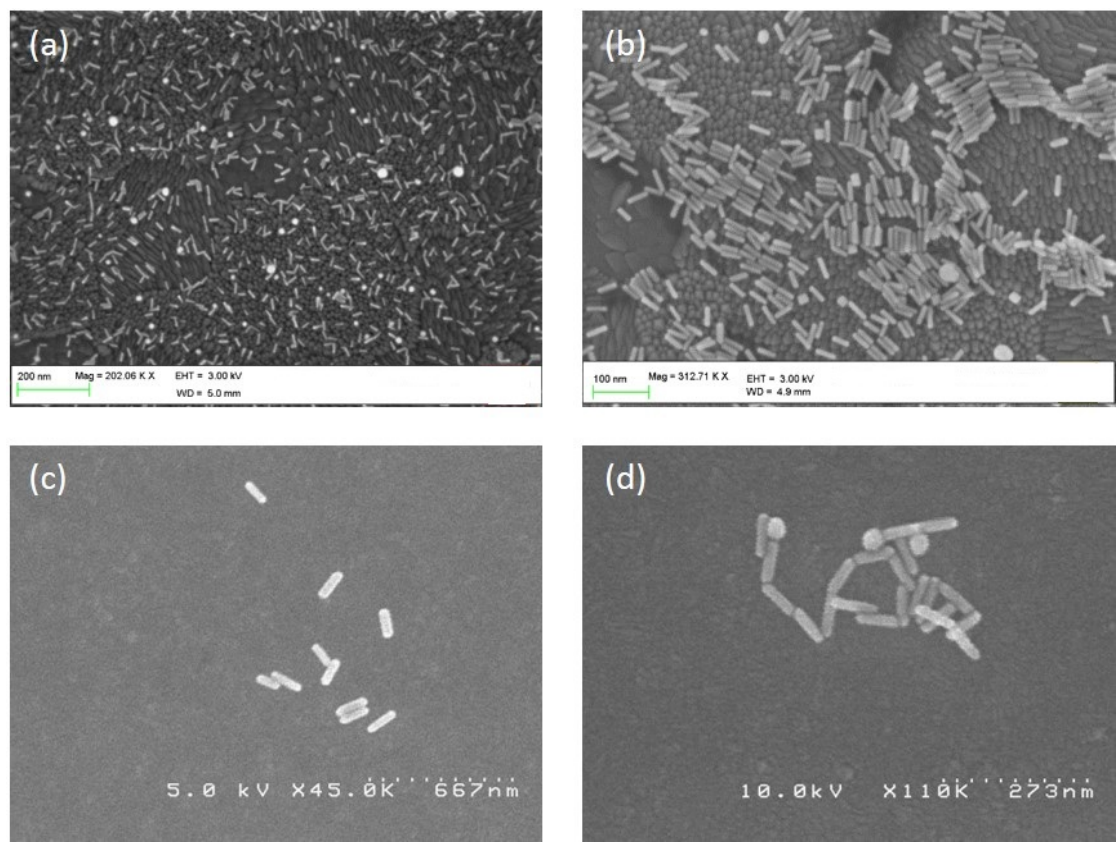

**Figure SP2** SEM Characterization of the different GNRs after their immobilization onto ITO coated glass cover slides (see Methods) : (a) GNR1 (d=6 nm, L= 27 nm) ; (b) GNR2 (d=10 nm, L =40 nm) ; (c) GNR3 (d=21.5 nm, L=84 nm, A12-25-808-CTAB) and (d) GNR4 (d=39 nm, L =145 nm, A12-40-808-CTAB). Although there appears a fairly narrow rod polydispersity, a few additional spheres can be observed, which is however not a problem since the morphology of each individual object is checked simultaneously as the TPL is measured. Moreover, sphere do not exhibit any TPL at the IR exciting wavelengths that have been considered in this work.

## 3 - Experimental set-up :

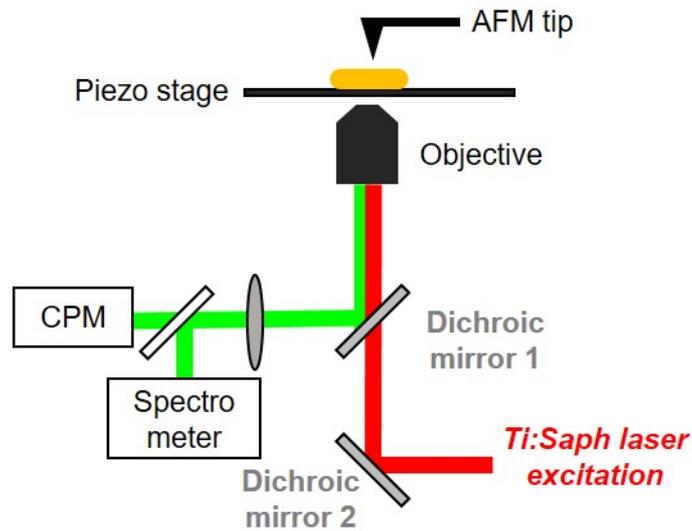

**Figure SP3:** Combined TPL and AFM microscopy set-up involving an inverted microscope (Olympus IX71) coupled to a cantilever type AFM platform (NanoWizard III, JPK) and associated to a femtosecond laser excitation. After alignment of the AFM tip with the focused laser spot, the sample is raster scanned enabling simultaneous topography (tapping mode) and TPL recording. Excitation is performed through a dichroic mirror (SemRock FF750-SDi02-25x36) and an oil immersion 100x microscope objective. The emitted TPL is collected through the same microscope objective and separated from the incident light by another dichroic mirror (SemRock FF735-Di670-25x36). The signal is then sent either to a channel plate multiplier working in the photon counting mode (Perkin Elmer MP-993-CL) or to a spectrometer coupled to a CCD camera (Andor DU401-BR-DD) for the detailed study of emission spectra.

## 4 - Polarization rate of the visible emission

|  | 10 nm x 40 nm | 21.5 nm x 40 nm | 39 nm x 145 nm |
|---|---|---|---|
| **Polarization rate** | 80 % | 70 % | 60 % |

**Figure SP4:** Table summarizing the polarization rate of the TPL emission in the visible for each considered GNRs batch.

## 5 – BEM simulations :

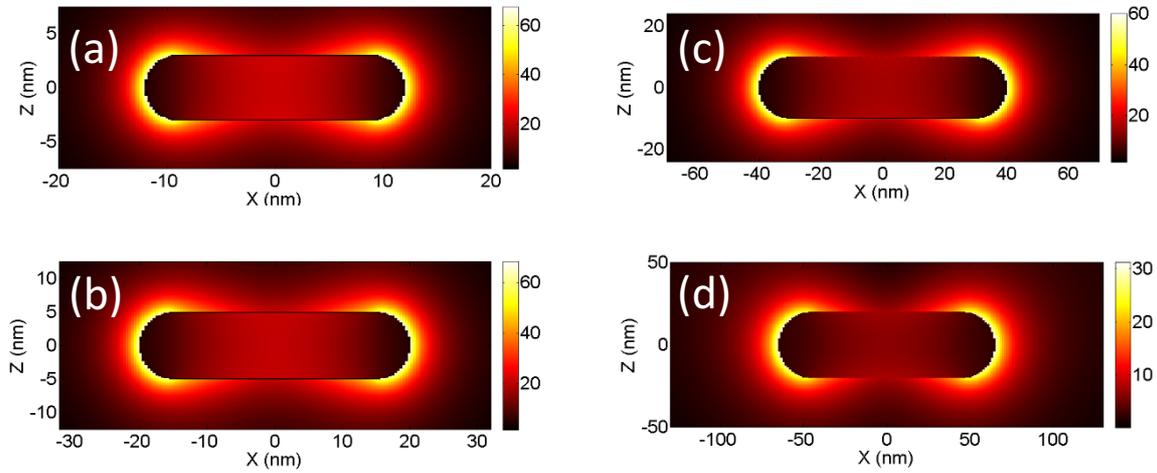

**Figure SP5:** Calculated electrostatic field enhancement map in intensity (*ie* $f^2_{loc,\text{exc}}$) for each GNR simulated with BEM : (a) 6 nm x 24 nm (b) 10 nm x 40 nm (c) 20 nm x 80 nm (d) 40 nm x 130 nm. The excitation wavelength is equal to the longitudinal resonance wavelength of each GNR (793 nm, 799 nm, 828 nm and 820 nm respectively) and the incident polarization is along the long axis of the GNR.

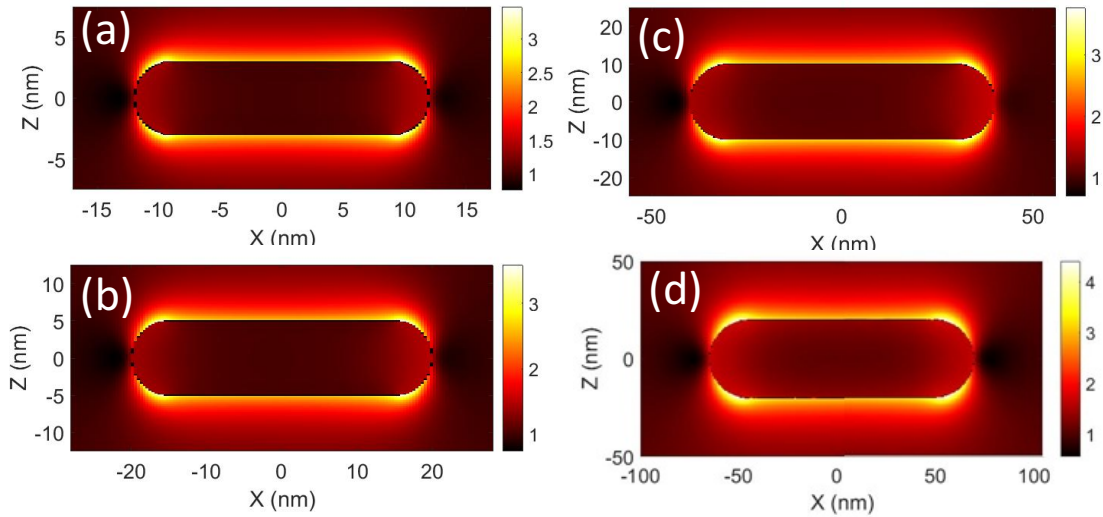

**Figure SP6:** Calculated electrostatic field enhancement map in intensity (*i.e.* $f^2_{loc,\text{TPL}}$) for each GNR simulated with BEM (a) 6 nm x 24 nm (b) 10 nm x 40 nm (c) 20 nm x 80 nm (d) 40 nm x 130 nm. The excitation wavelength is equal to the transverse maximum scattering wavelength of each GNR (522 nm, 522 nm, 522 nm and 534 nm respectively – see figure SP8) and the incident polarization is along the short axis of the GNR.

Before computing the field enhancement map for each GNR, their extinction, absorption and scattering cross section were calculated for a polarization along the longitudinal axis (figure SP7) and for a polarization along the short axis (figure SP8).

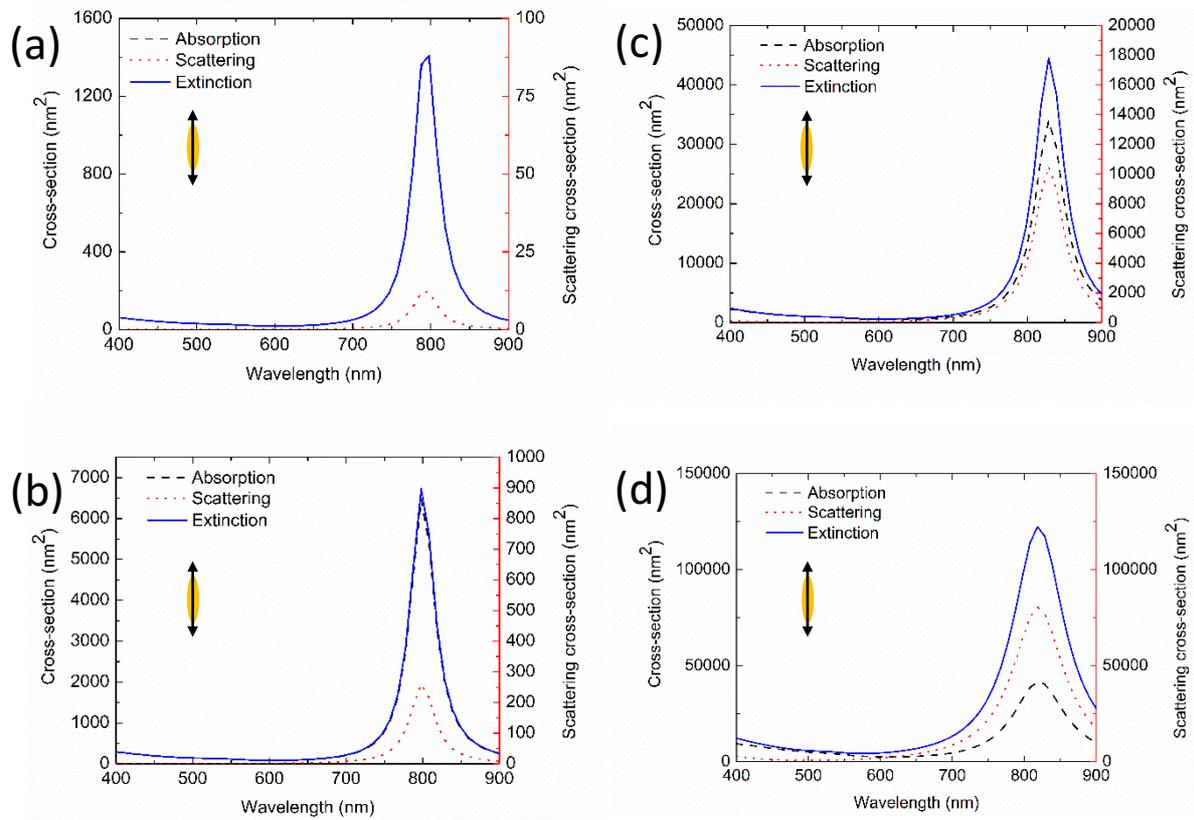

**Figure SP7:** Extinction (blue line), absorption (black dashed line) and scattering (red dotted line, corresponding to the right scale) cross-section calculated with BEM in retarded mode for each GNR (a) 6 nm x 24 nm (b) 10 nm x 40 nm (c) 20 nm x 80 nm (d) 40 nm x 130 nm for an incident polarization along the long axis of the GNR.

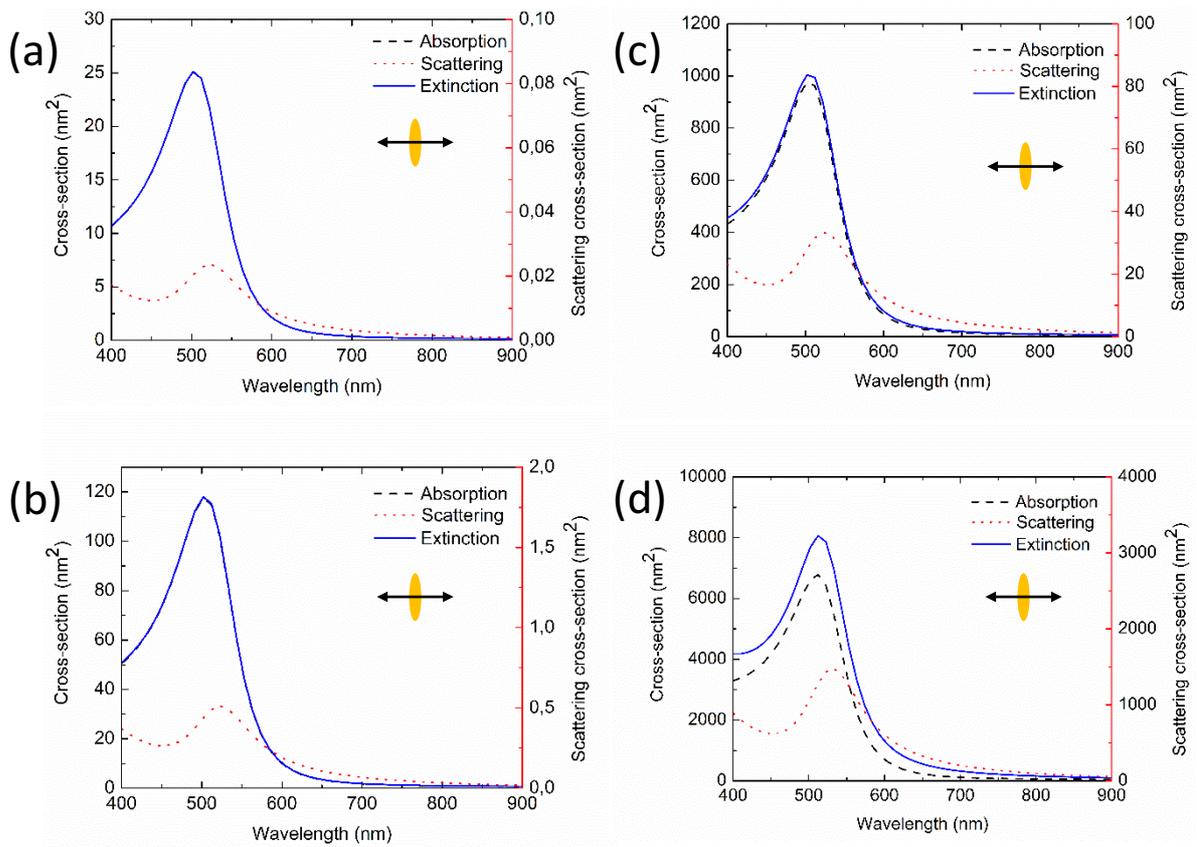

**Figure SP8:** Extinction (blue line), absorption (black dashed line) and scattering (red dotted line, corresponding to the right scale) cross-section calculated with BEM in retarded mode for each GNR (a) 6 nm x 24 nm (b) 10 nm x 40 nm (c) 20 nm x 80 nm (d) 40 nm x 130 nm for an incident polarization along the short axis of the GNR.

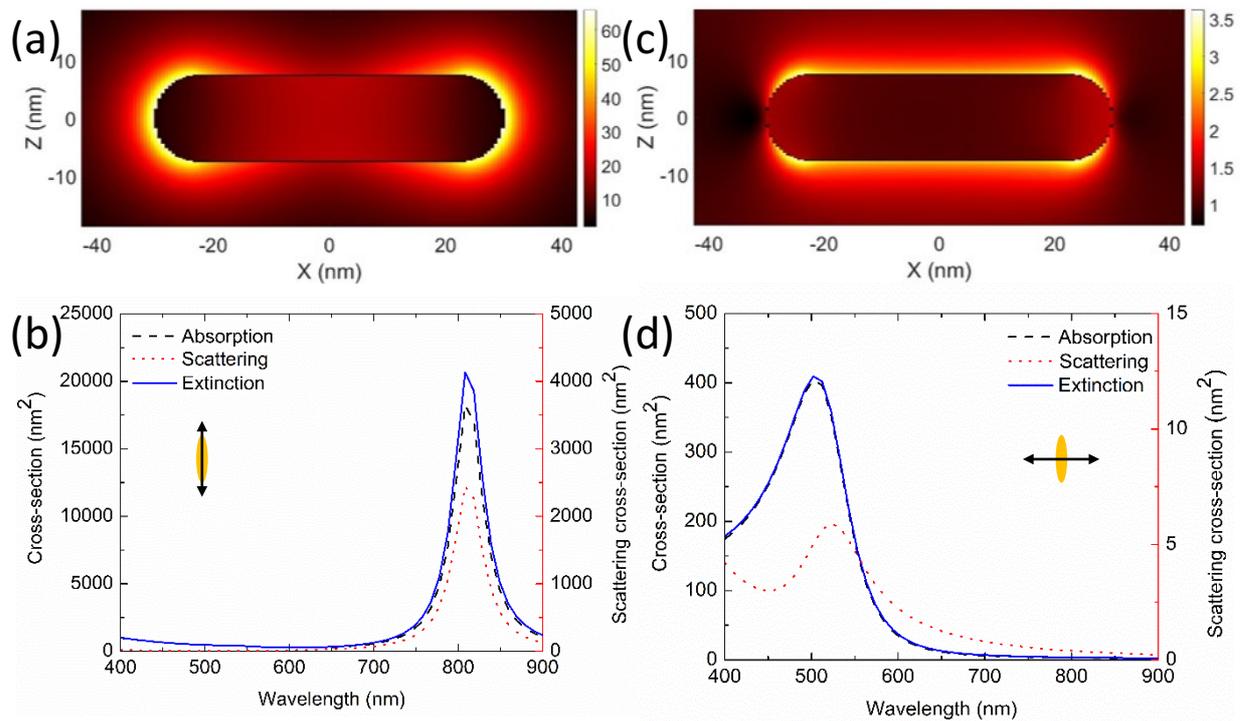

**Figure SP9:** BEM simulation of the electrostatic field enhancement map for a GNR with *d*=15 nm and *L*=60 nm in retarded mode, for an excitation wavelength equal to the longitudinal resonance wavelength of the GNR (808 nm) and an incident polarization along the GNR long axis (a) and for an excitation wavelength equal to maximum transverse scattering wavelength of the GNR (522 nm) and a polarization along the short axis of the GNR (c). Extinction (blue line), absorption (black dashed line) and scattering (red dotted line, corresponding to the right scale) cross-sections calculated with BEM in retarded mode for a 15 nm x 60 nm GNR for an incident polarization along the long (a) and short (b) axis of the GNR.